\documentclass[useAMS,usenatbib,usegraphicx]{mn2e}
\usepackage{times,hyperref,xcolor,amsmath,amssymb}
\pdfoutput=1

\newcommand\vect[1]{\ensuremath{\mathbf{#1}}}
\newcommand\kms{\ensuremath{\text{km~s}^{-1}}}
\newcommand\msun{\ensuremath{M_\odot}}
\newcommand\feh{\ensuremath{\mathrm{[Fe/H]}}}

\hypersetup{pdftitle=The Quest for the Sun's Siblings: an Exploratory Search in the Hipparcos Catalogue,
            pdfsubject=Paper on the search for the Sun's siblings in the
            Hipparcos Catalogue, pdfauthor=Anthony Brown - Simon Portegies Zwart - Jennifer Bean}

\begin{document}
%
\title[The Quest for the Sun's Siblings]{The Quest for the Sun's Siblings: an
Exploratory Search in the Hipparcos Catalogue}
\author[A.G.A. Brown, S.F. Portegies Zwart \& J. Bean]{Anthony G.A.\ Brown$^1$,
Simon F.\ Portegies Zwart$^1$ and Jennifer Bean$^2$\thanks{Euroscholar,
Sterrewacht Leiden, Leiden University}\\
$^1$Sterrewacht Leiden, Leiden University, P.O.\ Box 9513, 2300 RA, Leiden, The
Netherlands\\
$^2$Physics, Astronomy, and Materials Science Department, Missouri State
University, 901 S.\ National Avenue Springfield, MO 65897, USA}

\maketitle
\begin{abstract}
  We describe the results of a search for the remnants of the Sun's birth
  cluster among stars in the Hipparcos Catalogue. This search is based on the
  predicted phase space distribution of the Sun's siblings from simple
  simulations of the orbits of the cluster stars in a smooth Galactic potential.
  For stars within $100$ pc the simulations show that it is interesting to examine
  those that have small space motions relative to the Sun. From amongst the
  candidate siblings thus selected there are six stars with ages consistent with
  that of the Sun. Considering their radial velocities and abundances only one
  potential candidate, HIP 21158, remains but essentially the result of the
  search is negative. This is consistent with predictions by \cite{SPZ2009} on
  the number of siblings near the Sun. We discuss the steps that should be taken
  in anticipation of the data from the Gaia mission in order to conduct fruitful
  searches for the Sun's siblings in the future.
\end{abstract}


\section{Introduction}

The Sun's life history has long been a subject of interest not just in
astrophysics but also in fields such as solar system studies, the history of the
earth's climate, and understanding the causes of mass extinctions. The possible
birth environment of the Sun was discussed extensively by \cite{Adams2010} who
shows how inferences about this environment can be made by considering its
impact on the formation and morphology of our planetary system, the removal of
the solar nebula, and the presence of short-lived radioactive nuclei in
meteorites. The subsequent life and times of the Sun as it travels through our
Galaxy have attracted attention in the context of trying to understand climate
change and mass extinctions as the consequences of astronomical impacts. The
evidence for and against this idea was reviewed by \cite{BailerJones2009}, who
points out problems in the methodology of the various studies into climate
change or mass extinctions and also the uncertainties in the details of the
Sun's path through our Galaxy even over the past 545 Myr.

As discussed by \cite{SPZ2009} the Sun is likely to have been born in a bound
open cluster consisting of a few thousand stars. This cluster probably had a
radius of a few pc and as pointed out by \cite{Adams2010} the Sun was located
not too far from the cluster centre ($\sim0.2$ pc) as inferred from the
necessity of a nearby supernova explosion. The fact that the Sun thus has a
large `family' prompted \cite{SPZ2009} to ask the question: can we find the
Sun's siblings? The answer to this question is important as the inferences about
the Sun's birth environment all come from considering the Sun and its planets,
there is as yet no direct observational constraint on the birth cluster itself.
Identifying even a small number of the Sun's siblings would put constraints on
the number of stars in the cluster, by extrapolation for a plausible IMF, and
possibly even on the IMF itself if siblings were found over a range of stellar
masses. Reconstructing the orbits of the siblings in the Galaxy would lead to a
more accurate determination of the Sun's birth location as well as the
subsequent path to its present day position. This information could be used, for
example, to investigate whether the Sun's relatively high metallicity
\citep[cf.][]{Adams2010} can be explained by its birth at a different radius in
the Galaxy. In addition we would obtain a determination of the Sun's motion
through the Galaxy independent from the geological record, which was listed by
\cite{BailerJones2009} as an important goal for the study of the history of the
earth's climate and mass extinctions. \cite{SPZ2009} proceeded by considering
the constraints on the Sun's birth cluster and performing simple simulations of
the evolution of a cluster of stars initially confined to a $1$ pc virial radius
and orbiting our Galaxy along the presumable path the Sun followed in the past.
Depending on how quickly the cluster became unbound \cite{SPZ2009} concluded
that $\sim10$--$40$\% of the Sun's siblings should still be located with $1$ kpc
of the present day location of the Sun.

Thus we can expect to find about $\sim 100$--$1000$ of the Sun's siblings with
$1$ kpc from the present day position of the Sun. This will make a search for
the siblings extremely challenging as they will have to be weeded out from among
the $\sim 10^8$ stars within $1$ kpc. Nevertheless we set out in this paper to
make a first attempt at identifying candidate siblings of the Sun by searching
in the Hipparcos Catalogue and adding complementary data from the
Geneva-Copenhagen survey of the Solar neighbourhood \citep{GCS2009}. Our
motivation is to carry out a first exploration of kinematic searches for the
Sun's siblings. We describe our search methodology in section \ref{sec:method}
and present our results in section \ref{sec:results}. We discuss the results in
section \ref{sec:discussion} and outline the steps needed to carry out a
thorough future search for the Sun's siblings in section \ref{sec:future}.

\section{Search methodology}\label{sec:method}

As shown in \cite{SPZ2009} the Sun's siblings are expected to remain near the
Sun's orbital trajectory and will form a characteristic pattern in the proper
motion vs.\ distance plane. The specific distribution of the Sun's siblings in
phase space can thus be used to find candidate siblings in the Solar
neighbourhood. We pursue this idea by first simulating the orbits of the stars
in the Sun's birth cluster, starting from the presumed birth place of the Sun.
The latter is found by tracing back the Sun's orbit over 4600 Myr in an analytic
Galactic potential. The birth cluster is then generated and the orbits of all
the stars in the cluster integrated forward in time in order to find the present
day phase space distribution of the siblings that remain near the Sun.

The simulated phase space distribution can be used to make a first selection of
sibling candidates from the stars with known phase space data near the Sun. To
do this we will use the Hipparcos Catalogue \citep{ESA1997}, specifically the
re-reduced version of the catalogue by \cite{FVL2007}. This first selection has
to be narrowed down further by using the powerful constraints of the Sun's age
and metallicity as described in section~\ref{sec:resultsnarrow}.

\subsection{Simulations}\label{sec:simulations}

For this exploratory attempt at identifying candidate siblings of the Sun we
decided to keep the simulations of the Sun's birth cluster simple. The birth
cluster is simulated as a collection of stars with a Gaussian distribution in
position and velocity:
\begin{equation}
  \rho(r)\propto e^{{-r^2}/{2\sigma_r^2}}
  \quad\text{and}\quad
  f(v)\propto e^{{-v^2}/{2\sigma_v^2}}\,,
  \label{eq:clusterdef}
\end{equation}
where $r=|\vect{r}-\vect{r}_\odot(0)|$ and $v=|\vect{v}-\vect{v}_\odot(0)|$ are
the position and velocity of the cluster stars with respect to the Sun's
position and velocity at birth (at time $t=0$). The dispersions $\sigma_r$ and
$\sigma_v$ are in units of pc and \kms. The self-gravity of the cluster stars is
ignored in these simulations (the stars are all treated as test particles in the
Galactic potential) which amounts to assuming that the cluster rapidly disperses
after its formation. As shown by \cite{SPZ2009} this is actually the most
challenging case as then only about 10\% of the siblings are expected to be
found within $1$ kpc from the Sun at present. The clusters were simulated with
the following dispersions in position and velocity: $\sigma_r=1$, $3$ pc and
$\sigma_v=0.5$, $1$, $2$ \kms. These numbers span plausible birth cluster sizes
and velocity dispersions \citep[cf.][]{SPZ2009,Adams2010} and are consistent
with the assumed number of cluster stars. This can be shown by assuming a
\cite{Salpeter1955} IMF for a cluster of $3000$ stars with a mass range of
$0.1$--$50$ \msun, which implies a total cluster mass of $\sim1000$ \msun.  The
virial theorem can then be used to make a crude estimate of the velocity
dispersion of the cluster $\sigma_v^2\approx GM/\sigma_r$, which for cluster
masses of a few $100$ (for non-Salpeter IMFs) to $1000$ \msun\ leads to values
for $\sigma_v$ in the range $\sim0.7$--$2$ \kms. All combinations of the
$\sigma_r$ and $\sigma_v$ values listed above were used. Although the real birth
cluster is expected to have contained a few thousand stars, $10\,000$ stars were
simulated in each cluster in order to sample the present day phase space
distribution with sufficient resolution.

\begin{figure*}
  \includegraphics[width=0.45\textwidth]{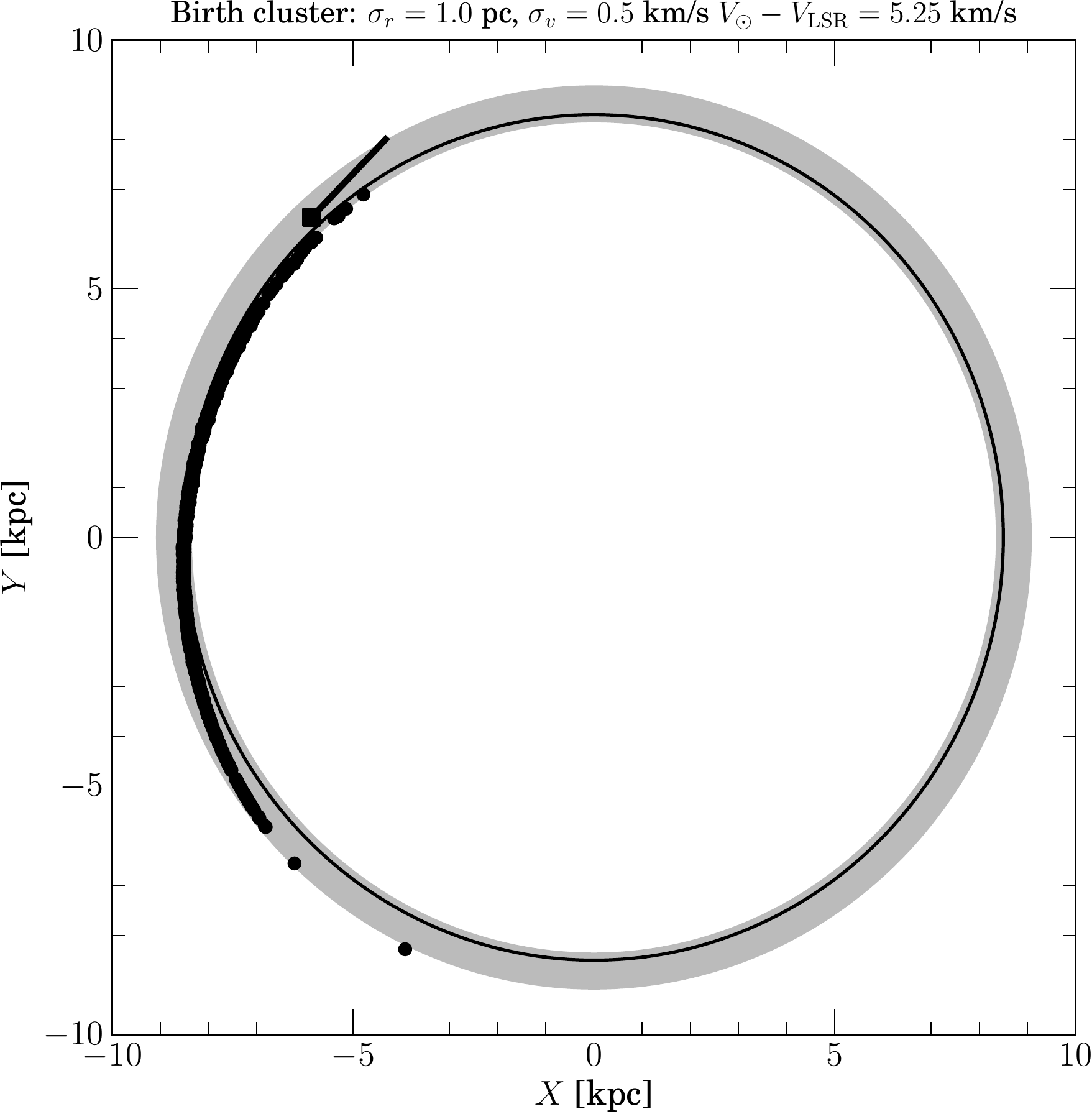}\hfil
  \includegraphics[width=0.45\textwidth]{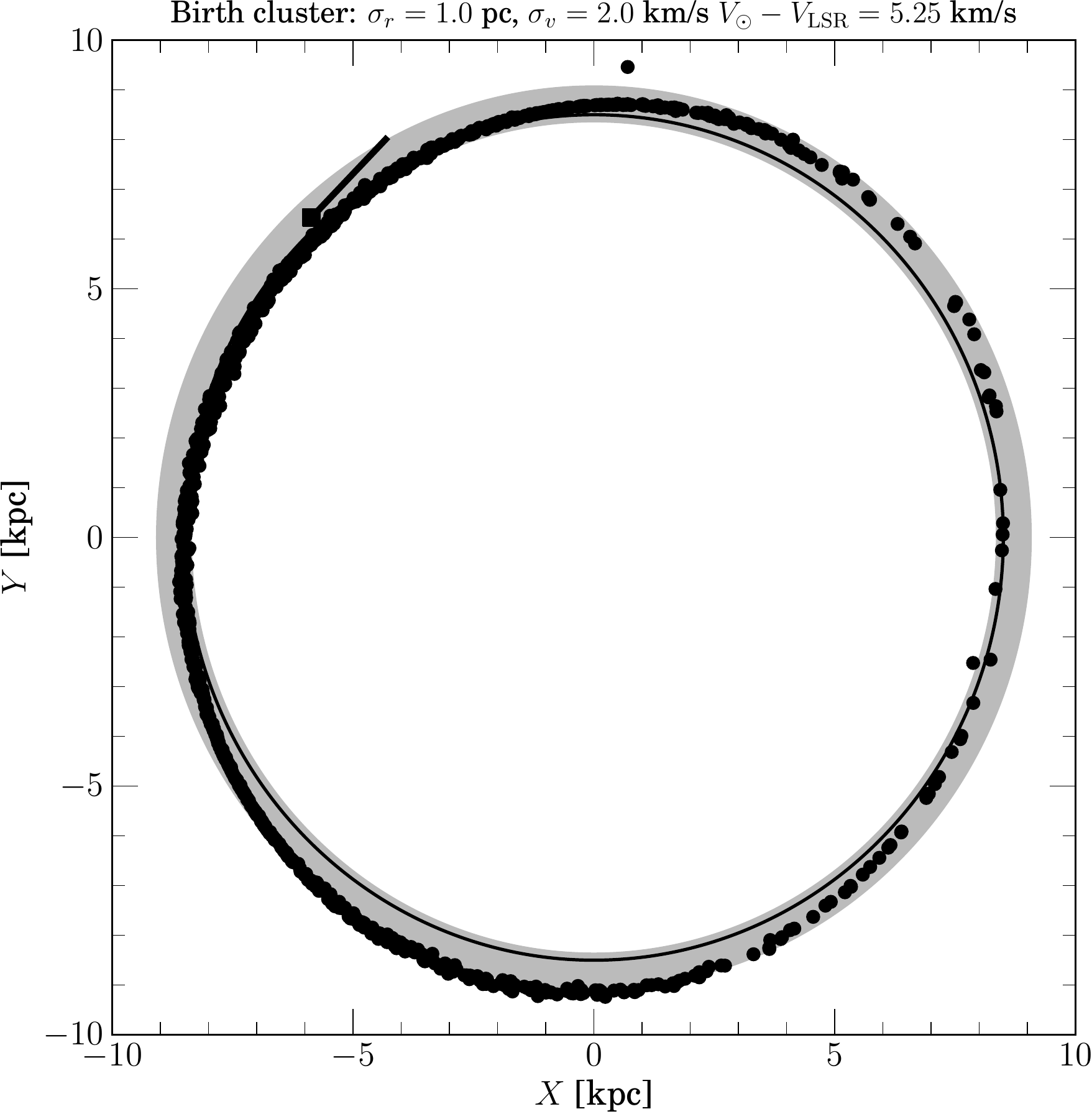}\\
  \includegraphics[width=0.45\textwidth]{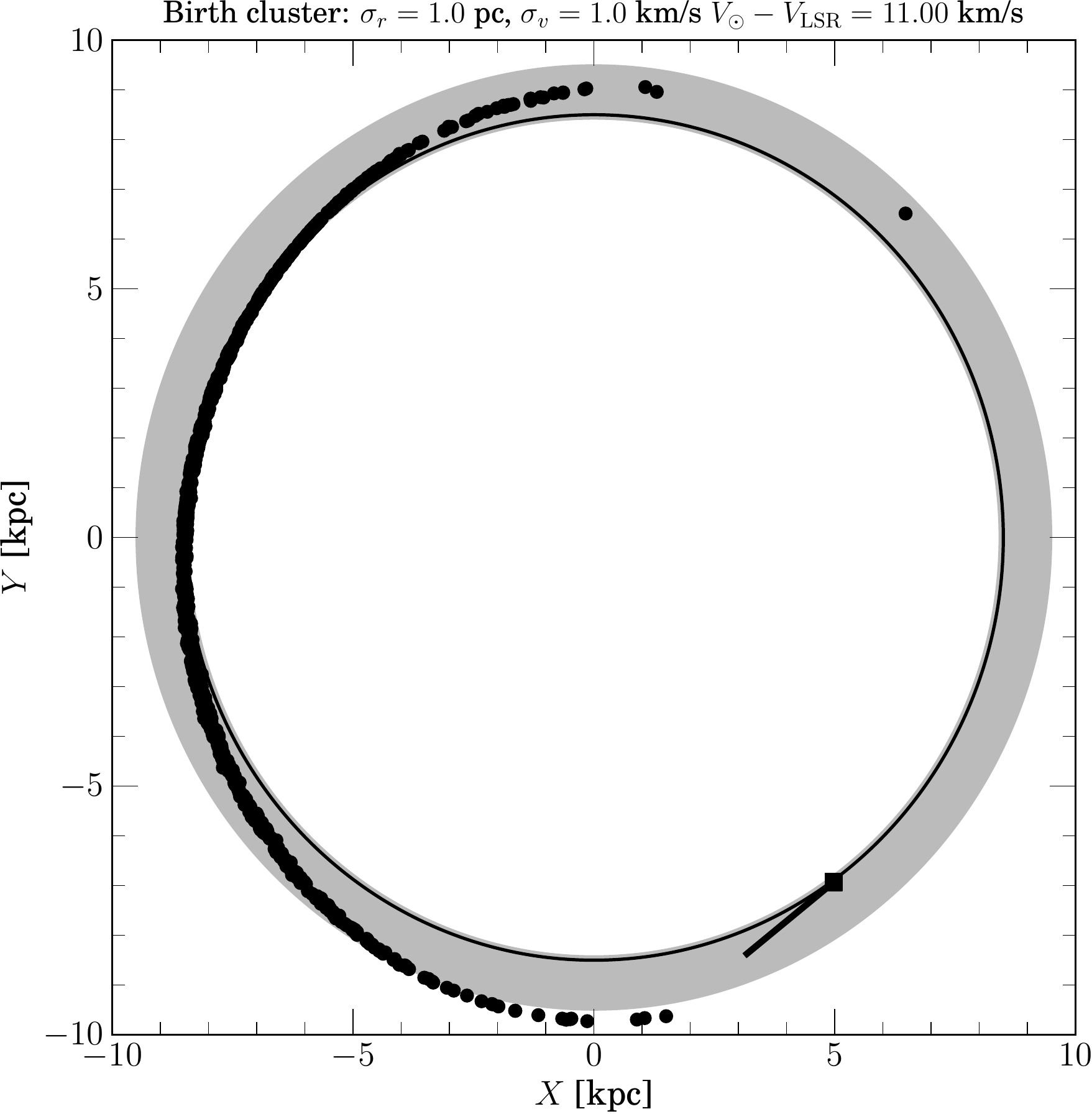}\hfil
  \includegraphics[width=0.45\textwidth]{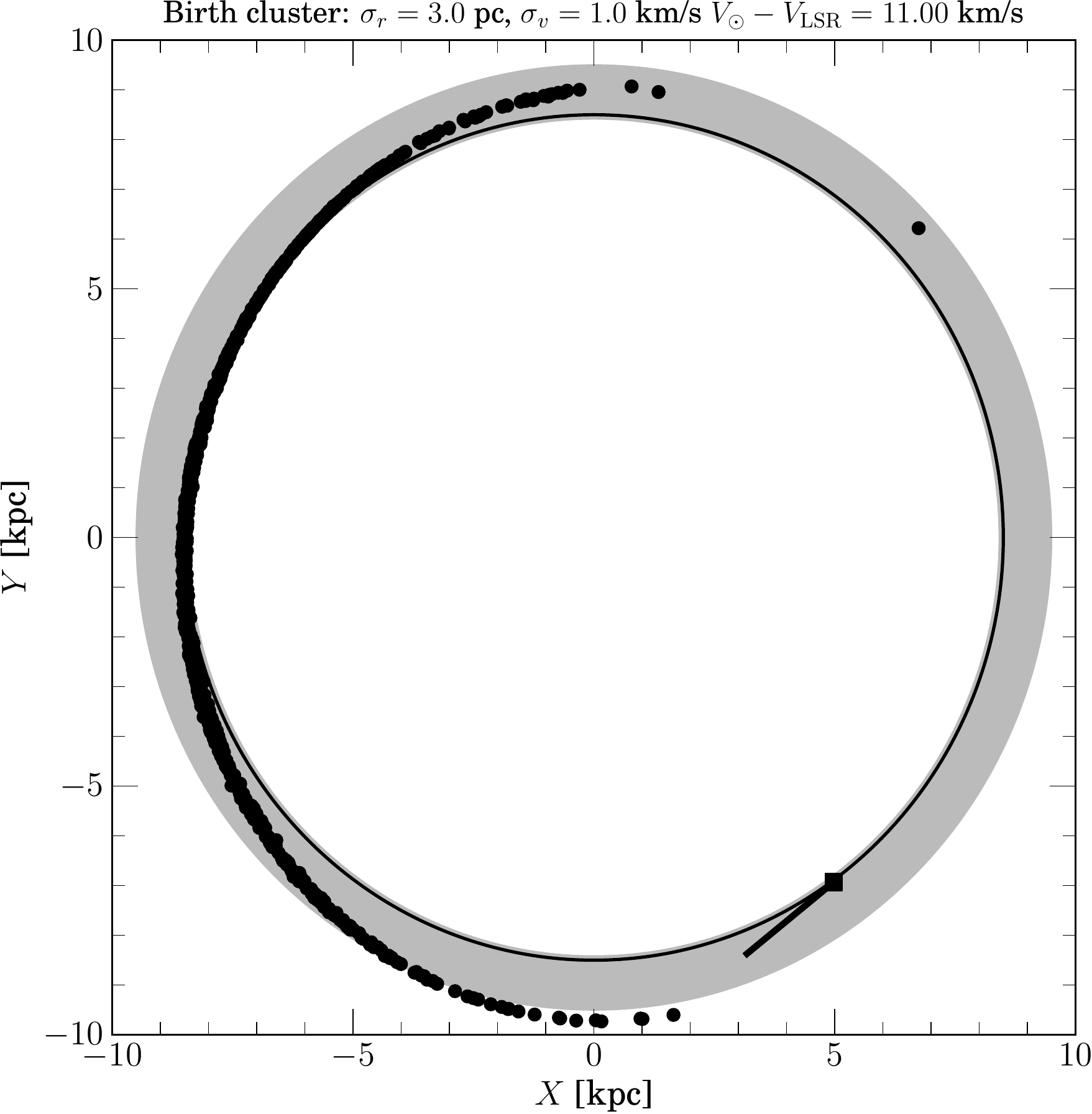}
  \caption{Distribution of the Sun's siblings projected on the Galactic plane
  for various birth cluster parameters. In all panels the grey ring indicates
  the radial extent of the Sun's orbit over the past 4600 Myr and the large dots
  show the present-day distribution of the siblings. Only one in every ten
  siblings is plotted. In the top two panels the present day value of $V_\odot$
  is $5.25$ \kms\ and in the bottom two panels the value is $11$ \kms.  Note the
  difference in the solar orbit and its birth position and velocity (indicated
  by the large square with the velocity vector attached). The top two panels
  illustrate the effect of increasing the birth cluster velocity dispersion from
  $0.5$ (left) to $2.0$ \kms\ (right); the siblings are spread much more along
  the Sun's orbit in the latter case. The bottom panels illustrate that the
  results are not sensitive to the size of the cluster which is $\sigma_r=1$ pc
  in the left panel and $\sigma_r=3$ pc in the right panel.  In all panels the
  large black circle indicates the orbit of the LSR in the \protect\cite{AS91}
  potential.\label{fig:xyprojections}}
\end{figure*}

The birth position of the Sun is found by integrating its orbit backward in
time, starting from the current position and velocity of the Sun. The present
day position of the Sun is fixed at $(X,Y,Z)_\odot=(-8.5,0,0)$~kpc in the
conventional Galactocentric Cartesian coordinate system.\footnote{Translated to
a Sun-centred reference frame the $X$-axis points toward the Galactic centre,
the $Y$-axis in the direction of Galactic rotation, and the $Z$-axis completes
the right-handed coordinate system.} The Sun's present day velocity with respect
to the local standard of rest was taken from \cite{Aumer2009} to be
$(U,V,W)_\odot=(9.96,5.25,7.07)$ \kms. Recently \cite{Binney2010} and
\cite{McMillan2010} advocated an upward revision of $V_\odot$ to $11$ \kms and
this value was also used to trace back the Sun's birth position. For both
starting positions in velocity space all the cluster configurations mentioned
above were simulated.

All orbit integrations were carried out in the analytic potential described in
\cite{AS91}, which consists of a Miyamoto-Nagai disk, a Plummer bulge, and a
spherical halo (in this potential the circular velocity at $R=8.5$~kpc is $220$
\kms). We used a 7th order Runge-Kutta integrator RK7(8) with the coefficients
listed in \cite{Fehlberg1967}. The integration time was fixed to 4600 Myr for
both the backward integration of the Sun and the subsequent forward integration
of the cluster stars. For the near-circular orbits integrated in these
simulations the energy conservation error is always at the machine precision
level.

Figure \ref{fig:xyprojections} shows the present-day distribution of the Sun's
siblings projected on the Galactic plane for four of the birth cluster
parameters mentioned above. The main message in this figure is that the
distribution of the Sun's siblings is mainly sensitive to the velocity
dispersion of the birth cluster. The initial size plays almost no role and the
value of the Sun's $V$ velocity mainly influences the birth position and
velocity but does not have an effect on the final distribution in space of the
siblings. The value of $V_\odot$ does have an effect on the velocity
distribution of the siblings as will be discussed below. From here on we will
only consider birth clusters with $\sigma_r=1$ pc.

\section{Selecting candidate siblings from the Hipparcos
Catalogue}\label{sec:results}

\begin{figure*}
  \includegraphics[width=0.45\textwidth]{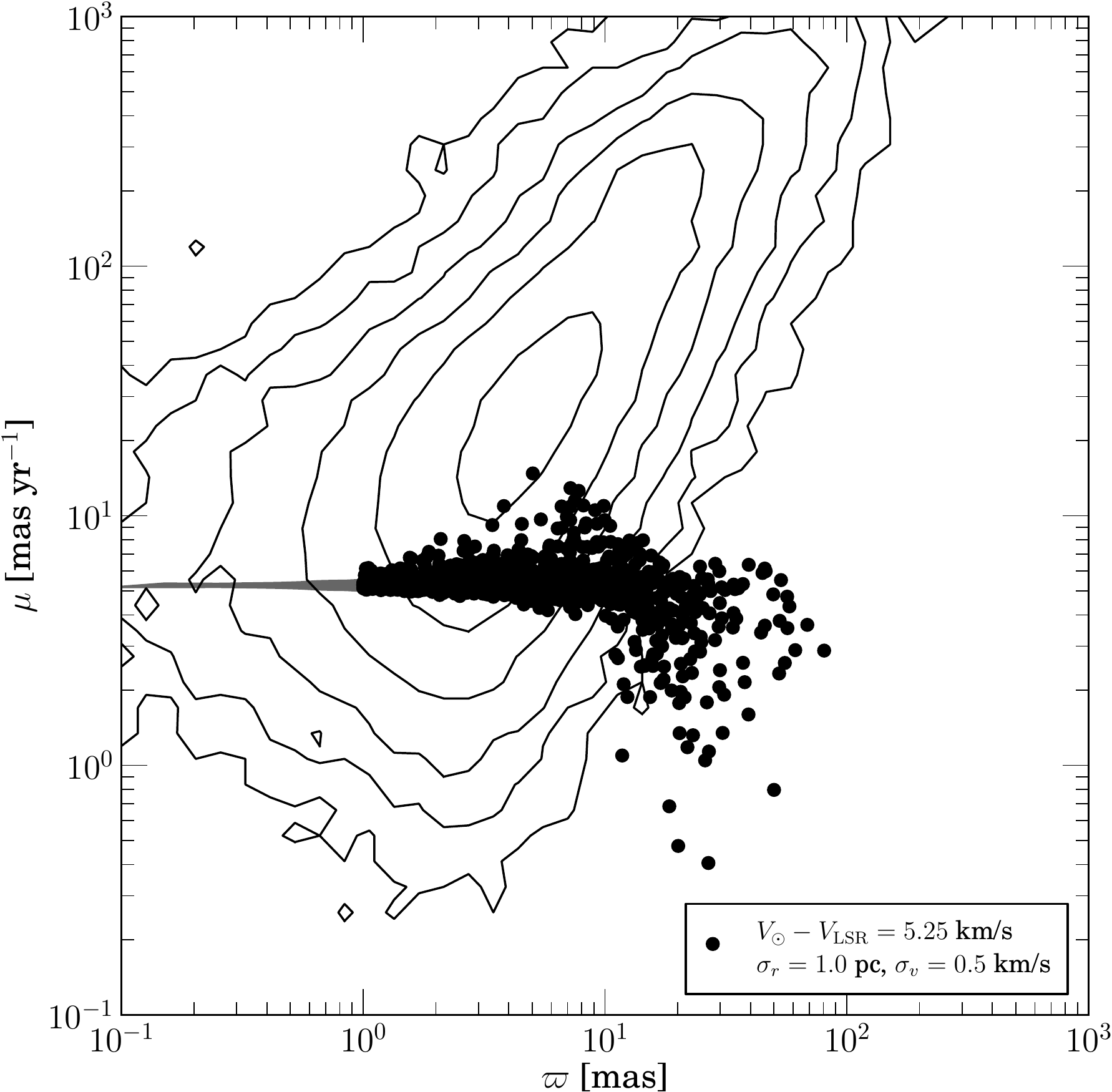}\hfil
  \includegraphics[width=0.45\textwidth]{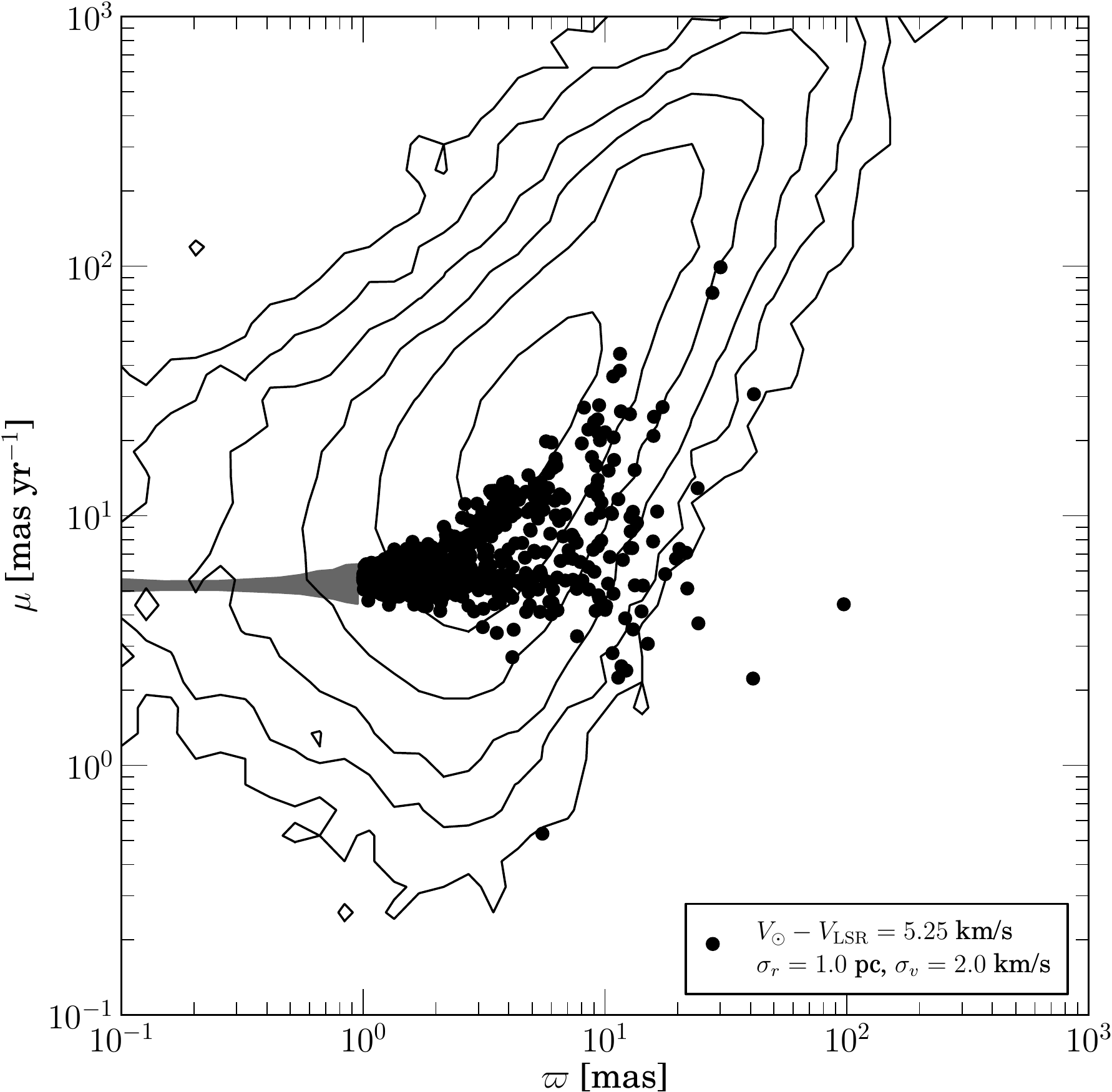}\\
  \includegraphics[width=0.45\textwidth]{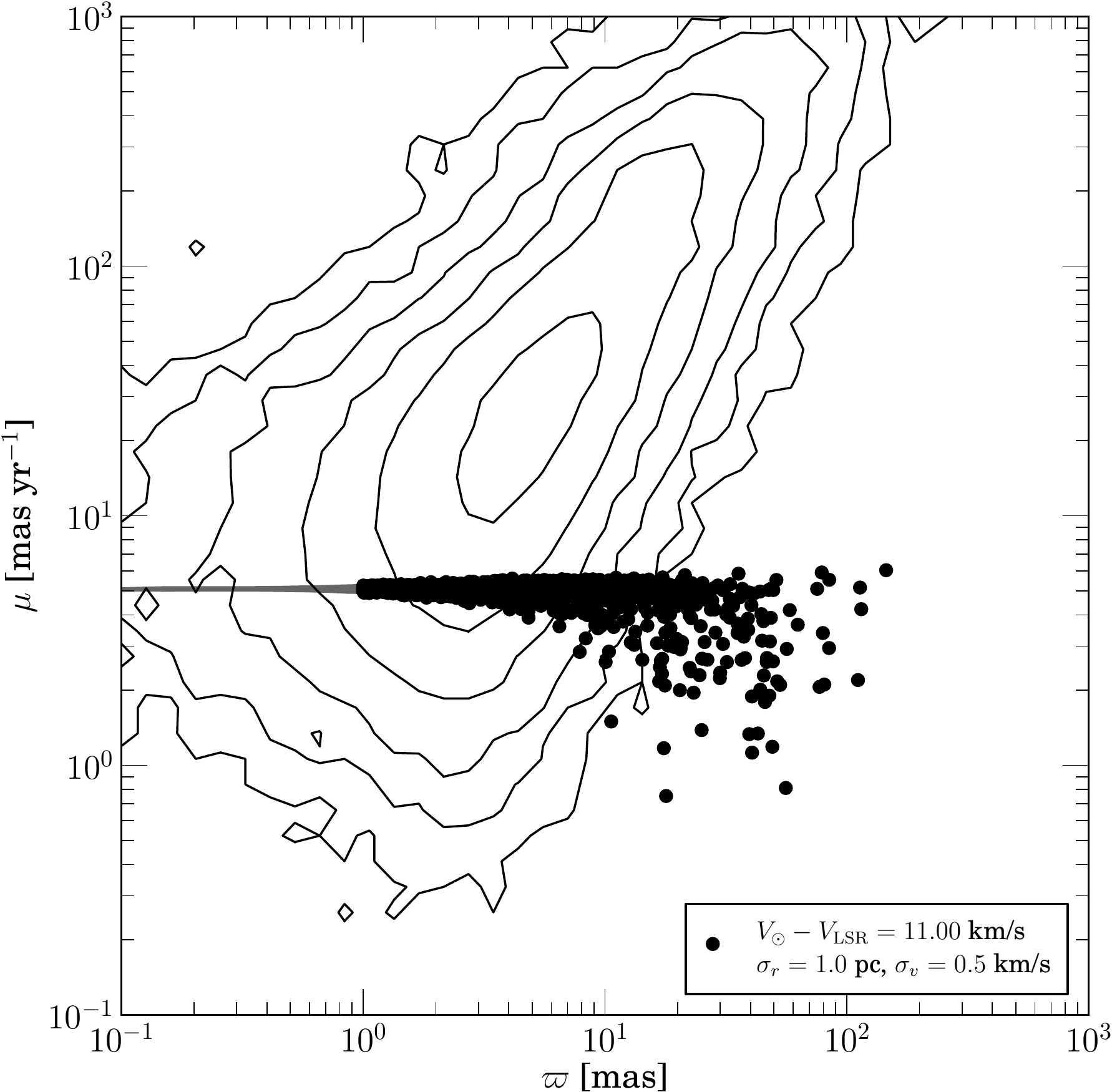}\hfil
  \includegraphics[width=0.45\textwidth]{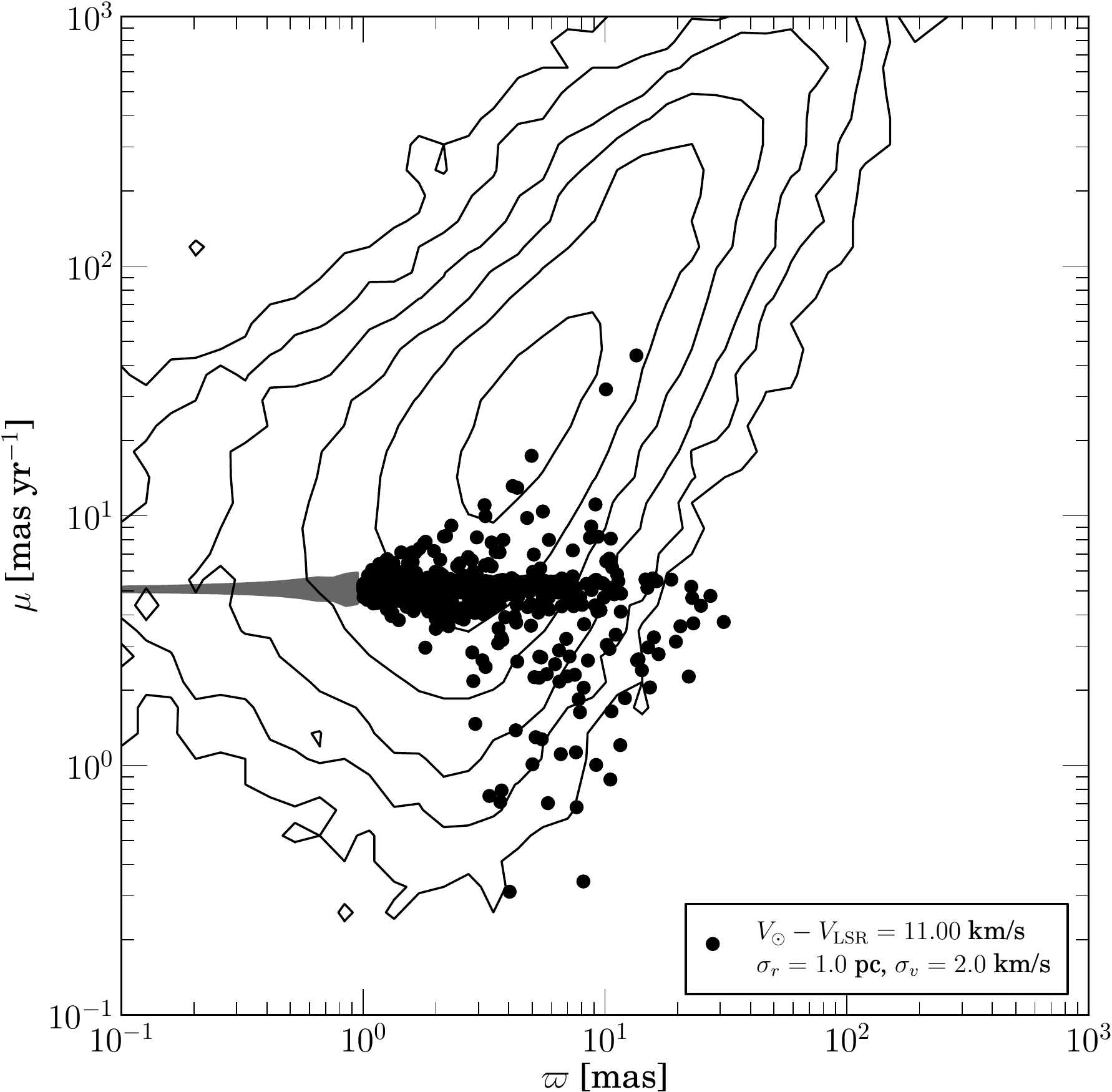}
  \caption{Distribution of the Sun's siblings in the proper motion ($\mu$) vs.\
  parallax ($\varpi$) plane. In all panels the distribution of Hipparcos
  Catalogue stars in this plane is shown as the contours, which indicate the
  numbers of stars in bins of $0.1\times0.1$ dex$^2$. The contour levels are at
  $3$, $10$, $30$, $100$, $300$, and $1000$ stars/bin. The dots show the
  simulated distribution of the siblings for low (left panels) and high velocity
  dispersion of the birth cluster, and for the low (top panels) and high (bottom
  panels) value of the $V$ component of the Sun's present day velocity. At
  $\varpi\leq1$ mas the distribution of the siblings is indicated with the grey
  region showing the mean proper motion $\pm3$ times the standard deviation in
  the distribution.\label{fig:pmparallax}}
\end{figure*}

As pointed out in the introduction the $\sim100$--$1000$ siblings within $1$ kpc
from the present day position of the Sun will have to be identified from among
the $\sim10^8$ mainly Galactic disk stars in the same volume. It will thus be
important to find a corner of phase space where there is a high contrast between
the siblings and the Galactic background. Following \cite{SPZ2009} we start by
examining the distribution of stars in the distance vs.\ proper motion plane,
which is shown in Fig.\ \ref{fig:pmparallax}. The contours in this figure show
the distribution of the Hipparcos Catalogue stars in proper motion vs.\
parallax. The overall shape of this distribution reflects the local Galactic
disk kinematics combined with the Hipparcos completeness limits. The overall
trend of proper motion with parallax is intrinsic to the Galactic disk. We
verified this by generating a mock all-sky catalogue of stars, complete to
$V=12$, using the Besan{\c c}on model \citep{Robin2003,Robin2004}.\footnote{The
model can be run at \texttt{http://model.obs-besancon.fr}.} The lack of stars at
proper motions below 1 mas/yr and parallaxes below a few mas is due to the
incompleteness of the Hipparcos Catalogue beyond $V\sim8$
\citep[cf.][]{ESA1997}. Important in this discussion is that the lack of stars
at low proper motion and high parallax is not caused by a selection bias.

We now examine the simulated distribution of the Sun's siblings in the proper
motion vs.\ parallax plane, indicated by the large dots in Fig.\
\ref{fig:pmparallax}. The simulated distributions are shown for low and high
velocity dispersion of the birth cluster and for the low and high values of the
$V$ velocity of the Sun. In all cases the proper motions converge to a value of
about $5$--$6$ mas/yr at small parallaxes (large distance). This can be
understood by considering stars that are moving along the solar circle with a
velocity comparable to that of the Sun. Their proper motion will at large
distance converge to $\sim V_\mathrm{LSR}/(4.74R_\odot)$. In fact at large
distances the upper and lower limit on the proper motion are given roughly by
$\mu\sim (V_\mathrm{LSR}\pm V_\odot)/(4.74R_\odot)$, where $R_\odot$ is the
distance from the Sun to the Galactic centre and $V_\mathrm{LSR}$ is the
velocity of the local standard of rest. The ratio
$(V_\mathrm{LSR}+V_\odot)/R_\odot$ was constrained by \cite{McMillan2010}, from
observations of Galactic maser distances and motions, to lie in the range
$29.8$--$31.5$ \kms kpc$^{-1}$, which expressed as a proper motion is
$6.3$--$6.6$ mas/yr. At close distances (large parallaxes) the siblings can
exhibit both higher and lower proper motions, where the upper and lower limits
on the proper motion value can again be roughly obtained by considering stars on
the same (nearly circular) orbit as the Sun and taking the varying distance into
account. Proper motions larger than $\sim5$--$6$ mas/yr only occur if $\sigma_v$
is relatively large compared to $V_\odot$ ($\sigma_v/V_\odot\gtrsim 0.1$ judging
from Fig.\ \ref{fig:pmparallax}).  However, in all cases shown in Fig.\
\ref{fig:pmparallax} there is a group of siblings at low proper motion and high
parallax, occupying the part of the diagram where few disk stars are expected.
As a first selection of candidate siblings of the Sun we therefore choose the
sample of Hipparcos stars with:
\begin{equation}
  \varpi\geq 10\text{ mas}\,\wedge\,
  \sigma_\varpi/\varpi\leq0.1\,\wedge\,
  \mu\leq6.5\text{ mas/yr}\,,
  \label{eq:candsibl}
\end{equation}
where we additionally select on the parallax precision. We note that the
siblings with these characteristics are predicted to have radial velocities of
less than $\sim10$ \kms\ in absolute value, where the distribution is rather
strongly peaked around $v_\mathrm{rad}\sim0$ \kms. In our selection
(\ref{eq:candsibl}) we make use of the observationally established value of
$(V_\mathrm{LSR}+V_\odot)/R_\odot$ in order to avoid introducing biases related
to inadequacies in the simulated phase space distribution of the siblings.

This first selection of candidate siblings is mainly a quantitative statement of
the search for nearby stars on almost the same orbit as the Sun. The number of
candidate siblings after this first cut is $87$. In the following section we
further examine these stars by cross matching them against the Geneva-Copenhagen
survey \citep{GCS2009} and considering their ages and metallicities.

\subsection{Narrowing down the candidate list}\label{sec:resultsnarrow}

\begin{figure}
  \includegraphics[width=0.5\textwidth]{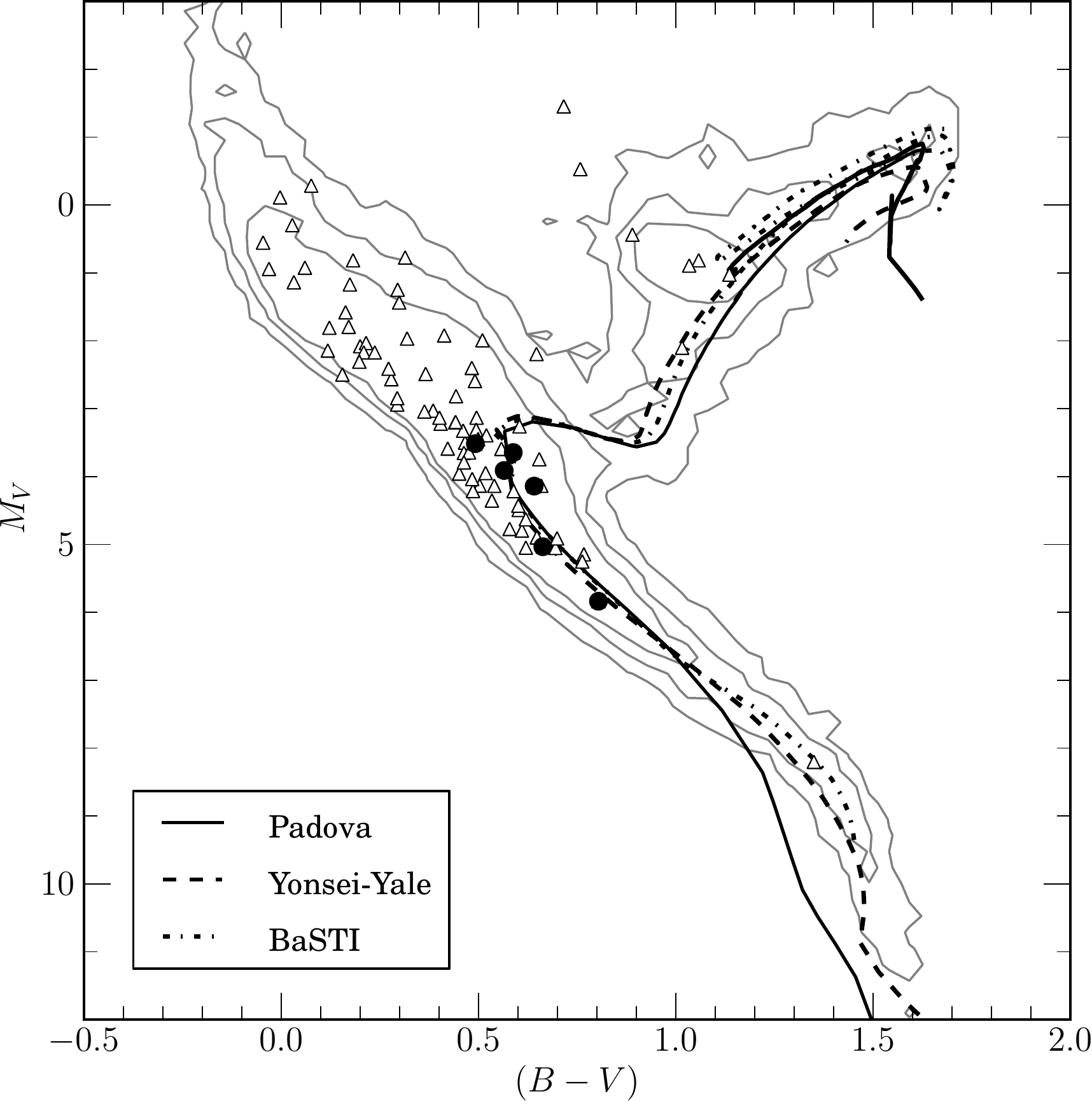}
  \caption{Colour magnitude diagram showing the absolute magnitude $M_V$ vs.\
  $(B-V)$. The contours show the distribution in this diagram of the stars in
  the Hipparcos Catalogue with $\sigma_\varpi/\varpi\leq0.1$ and
  $\sigma_{(B-V)}\leq0.05$ \protect\citep[data from][]{FVL2007}. The contours
  show the numbers of stars in bins of $0.05\times0.2$ mag$^2$, where the
  contour levels are at $5$, $20$, $50$, and $500$ stars/bin. The triangles are the
  candidate siblings selected according to (\ref{eq:candsibl}) and the large
  dots are the siblings selected from \protect\cite{GCS2009} with ages
  consistent with that of the Sun ($4.6$~Gyr). The solid line shows the
  isochrone at the age and metallicity of the Sun according to the Padova models
  \protect\citep{Marigo2008}, the dashed line shows the same isochrone for the
  Yonsei-Yale models \protect\citep{YY2004}, and the dot-dashed line for the
  BaSTI \protect\citep{BaSTI2004} models.\label{fig:cmd}}
\end{figure}

In figure \ref{fig:cmd} we show the colour magnitude diagram for the candidate
siblings selected according to (\ref{eq:candsibl}). The absolute magnitudes
$M_V$ were calculated using the $V$-band magnitudes from the Hipparcos Catalogue
\citep{ESA1997} and the parallaxes from the new reduction \citep{FVL2007}. Only
stars with precise parallaxes ($\sigma_\varpi/\varpi\leq0.1$) and colours
($\sigma_{(B-V)}\leq0.05$) were selected to produce the contours. The triangles
in figure \ref{fig:cmd} are the candidate siblings selected according to
(\ref{eq:candsibl}). The figure also shows three isochrones at the age,
$4.6$~Gyr, and composition of the Sun according to the Padova \citep[solid
line,][]{Marigo2008}, Yonsei-Yale \citep[dashed line,][]{YY2004}, and BaSTI
\citep[dot-dashed line,][]{BaSTI2004} stellar models.

From the location of the isochrones it is clear that we can (not surprisingly)
exclude stars with $(B-V)<0.5$ as candidate siblings of the Sun, they are simply
too young. Similarly the three giant stars at $0.5<(B-V)<1.0$ and $M_V<+1$ can
be excluded as candidate siblings. The rest of the stars cannot be excluded as
candidate siblings {\em on the basis of the information used so far}.

To further narrow down the list of candidate siblings we need to examine the
ages and compositions of the remaining stars, ideally using spectroscopy to
determine the astrophysical parameters of the stars and their chemical
compositions. However as a first step we cross-correlated the list of Hipparcos
selected candidate siblings against the data from the Geneva-Copenhagen survey
\citep{GCS2009,GCS2007,GCS2004}. We made use of the age estimates in this survey
to make a further selection among the candidate siblings by demanding that the
age is consistent with $4.6$ Gyr to within the confidence limits quoted in
\cite{GCS2009}. This results in the 6 candidate siblings indicated by the dots
in figure \ref{fig:cmd}. We list the properties of these stars in Table
\ref{tab:siblings}. The masses for these stars are in the range $0.8$--$1.2$
solar masses according to \cite{GCS2004}.

\begin{table*}
  \begin{minipage}{140mm}
    \caption{Observational properties of the candidate solar siblings selected
    from the Hipparcos and Geneva-Copenhagen Survey catalogues. The value of
    \feh\ and the ages are from \protect\cite{GCS2009}. The radial velocity was taken
    from \protect\cite{GCS2004}.\label{tab:siblings}}
  \centering
    \begin{tabular}{crrrrrrr}
      \hline
      HIP & \feh & clAge\footnote{Lower confidence limit on age.} & Age &
      chAge\footnote{Upper confidence limit on age.} &
      $(B-V)$ & $M_V$ & $v_\mathrm{rad}$\\
      & & Gyr & Gyr & Gyr & & & \kms\\
      \hline
      21158 & $ 0.04$ & $4.1$ & $5.3$ &  $7.0$ & $0.64$ & $4.14$ &  $6.6$ \\
      30344 & $-0.05$ &       & $1.4$ &  $7.6$ & $0.66$ & $5.03$ & $14.4$ \\
      51581 & $ 0.00$ & $3.4$ & $3.8$ &  $5.4$ & $0.59$ & $3.65$ & $17.4$ \\
      80124 & $-0.27$ & $3.6$ & $4.2$ &  $6.0$ & $0.56$ & $3.91$ & $-2.1$ \\
      90112 & $-0.19$ &       & $1.2$ & $16.1$ & $0.80$ & $5.84$ & $26.1$ \\
      99689 & $-0.27$ & $3.2$ & $3.6$ &  $4.6$ & $0.49$ & $3.52$ & $-4.4$ \\
      \hline
    \end{tabular}
  \end{minipage}
\end{table*}

For two of the selected stars (HIP 30344 and 90112) the most likely ages are of
the order of $1$~Gyr and both of them have relatively high radial velocities,
making it unlikely that they are siblings of the Sun. Of the four stars with
ages similar to the Sun (all located near the turn-off point on the isochrones)
there are two (80124 and 99689) that have somewhat low values of \feh\ and one
(51581) which has $\feh=0$ but a  rather high radial velocity. This leaves the
star HIP 21158 as the most likely candidate to be a sibling of the Sun. However
with a parallax of $26$ mas its radial velocity is high compared to the
predicted radial velocity of $\lesssim2$ \kms.

We note that the 5 stars at $(B-V)>1.0$ selected as candidate siblings from the
Hipparcos Catalogue all lie near the solar age isochrone. These are HIP 56287,
57791, 89825, 92831, and 101911, of which none occurs in the Geneva-Copenhagen
survey. The SIMBAD\footnote{http://simbad.u-strasbg.fr/simbad/} database was
used to find the values of \feh\ and $v_\mathrm{rad}$ for these stars. The first
three are unlikely to be siblings on account of their radial velocities which
are all larger than $10$ \kms. The metallicity ($\feh=-0.10$) and radial
velocity ($-8.7$ \kms) for HIP 92831 ($(B-V)=1.03$, $M_V=0.9$) are consistent
with being a sibling but no determination of its age exists. For HIP 101911
($(B-V)=1.02$, $M_V=2.1$) no information was found on its radial velocity or
\feh\ value.

\section{Discussion}\label{sec:discussion}

Our search for siblings of the Sun in the Hipparcos Catalogue thus leaves us
with at most 1 candidate sibling for which the radial velocity is on the high
side (at least if the simulations from section \ref{sec:simulations} are to be
trusted, see below). This is consistent with the fact that within $100$ pc from
the Sun only $0.1$--$1$ sibling is expected according to the simulations done by
\cite{SPZ2009} for plausible numbers of stars in the Sun's birth cluster
\citep{SPZ2009,Adams2010}.

For any reasonable cluster IMF it should be much more likely to find siblings of
lower mass than the Sun rather than the slightly more massive candidates from
table \ref{tab:siblings}. However, our search for siblings has been based on
data which is incomplete even for the nearest $100$ pc to the Sun. This is mainly
caused by the need for accurate trigonometric parallaxes which are used to place
stars precisely in the colour-magnitude diagram before determining their ages.
The Hipparcos catalogue is complete only to $V\sim7$--$8$ \citep{Turon1992}
which biases the sample of stars to those with masses above $\sim 1$\msun. This
selection effect is made stronger by the choice of sample for the
Geneva-Copenhagen survey \citep[cf.][]{GCS2004}. The effect is to bias our
present search to stars that are brighter than the Sun. Of these stars the most
numerous will be the ones near the main-sequence turn-off region for the solar
age isochrone, which is where most of the candidates in table \ref{tab:siblings}
are found.

We have restricted our search to the very nearby stars (within $100$~pc) because
of the higher contrast between possible siblings and field stars but also
because the Geneva-Copenhagen survey is restricted to the nearest $40$ pc around
the Sun. This survey provides the most comprehensive (and readily available) set
of consistent ages and metallicities for stars near the Sun and thus forms an
important source of information in this study. Many more candidate siblings can
be found at larger distances from the Sun and we did attempt to apply a
different selection of sibling candidates. Siblings at distances between
$\sim100$ pc and $\sim 1$ kpc from the Sun are predicted in our simulations to
cluster on the sky around the positions $(\ell,b)=(90^\circ,0^\circ)$ and
$(270^\circ,0^\circ)$, which can be appreciated by examining figure
\ref{fig:xyprojections}. We thus selected candidate sibling in these regions of
the sky from the Hipparcos catalogue and then further restricted the sample by
selecting on parallax and proper motion using figure \ref{fig:pmparallax}.  This
resulted in many candidates beyond the reach of the Geneva-Copenhagen survey and
did not turn up additional candidates.  Making use of other surveys that provide
astrophysical information from spectroscopy, such as RAVE
\citep{RAVE2006,RAVE2008} or SEGUE \citep{SEGUE2009} is likely to be
unsuccessful. In the case of RAVE the Galactic plane is not sufficiently covered
and in the case of SEGUE the targets are too faint to appear in the Hipparcos
Catalogue.

\section{Conclusions and future work}\label{sec:future}

Motivated by the desire to find the remnants of the Sun's birth cluster we have
conducted a preliminary search in the Hipparcos Catalogue for stars that could
have been born in the same cluster. This search was based on the predicted phase
space distribution of the Sun's siblings from simple simulations of the orbits
of the cluster stars in a smooth Galactic potential. For nearby stars the
simulations show that it is interesting to examine those that have small space
motions relative to the Sun. From amongst the candidate siblings thus selected
there are six stars with ages consistent with that of the Sun. Of these six
candidate siblings 5 can be excluded on the basis of their radial velocity or
metallicity, leaving only one plausible candidate sibling, HIP 21158.  However,
the latter still has a radial velocity somewhat higher than predicted from our
simulations. This means we have not found a single convincing solar sibling
within $100$ pc from the Sun which is consistent with the predictions by
\cite{SPZ2009} and the fact that only a small fraction of the stars near the Sun
was examined.

Now, even if a stronger case could have been made for the candidate siblings in
table \ref{tab:siblings} based on their age and value of \feh, this would not
have proven that these stars are truly siblings of the Sun. We discuss below
what steps need to be taken in future searches for the Sun's siblings.

The process of cluster disruption in the Galactic potential was simulated in a
simplified manner in both this work and in \cite{SPZ2009}. A better
understanding of the expected distribution of the siblings in phase space is
essential for an efficient search for them in future large surveys. Hence it is
important to do simulations of cluster disruption that are as realistic as
possible. Effects to be included are the self-gravity of the cluster stars,
non-axisymmetric structures in the Galactic potential, such as the bar and
spiral arms, and the collisions of the cluster with molecular clouds in the
Galaxy. The resulting phase space distribution is expected to be less orderly
than depicted in figures \ref{fig:xyprojections} and \ref{fig:pmparallax} but to
what extent is unknown at the moment. In these simulations it will be important
to ensure that the resolution is comparable to realistic cluster and molecular
cloud mass scales. To properly understand selection effects in surveys it is
also necessary to include a realistic initial mass function and stellar
evolution in the cluster simulations.

The observational challenge is equally daunting. On the one hand a large scale
survey of phase space is needed, covering a large volume of the Galactic disk.
Only the Gaia mission \citep{Gaia2008} will provide this data at the precision
needed to probe for siblings far away from the Sun. The above simulations will
have to be exploited to develop efficient search methods that can weed out the
candidate siblings from among the billion stars in the Gaia catalogue. However a
search in phase space only is not sufficient as the stars from different
clusters on similar orbits as the Sun's birth cluster could be confused with the
genuine siblings. In addition it is known that clustering of stars in phase
space can also be caused by dynamical effects \citep[see for
example,][]{Antoja2009}. The phase space search will have to be complemented by
a very detailed astrophysical characterization of candidate siblings of the Sun.
The age and overall abundance of the stars will not be enough in this respect as
many clusters with abundances similar to the Sun's birth cluster may have formed
around the same time. In addition the errors on individual stellar ages
($\sim0.5$--$1$ Gyr in table \ref{tab:siblings}) are likely to remain larger
than any plausible age spread within the birth cluster or the lifetime of the
molecular cloud from which the cluster formed. However, the Sun's siblings are
expected to have the same detailed chemical composition as the Sun and true
siblings can thus be identified through the analysis of high resolution spectra.
The latter will have to be collected in a dedicated follow-up programme.

This `chemical tagging' of stars has been proposed as a powerful method for
associating them with their formation sites \citep{Freeman2002}. So far it has
been demonstrated for the Hyades, the HR1614 moving group, and Cr 261 that these
groups of stars indeed have unique chemical signatures and promising elements
have been identified that can be used to chemically identify groups of stars
\citep{DeSilva2007}. However, a number of important studies regarding this
technique remain to be done:
\begin{itemize}
  \item No attempt has been made so far to {\em identify} (new) moving groups or
    clusters on the basis of abundance patterns, so the feasibility of this
    important aspect of the chemical tagging method remains to be demonstrated.
  \item It has not yet been definitively established to what accuracy the
    abundance patterns of stars have to be measured in order to identify them
    with their birth sites. Although \cite{DeSilva2007} conclude that $\sim0.05$
    dex accuracy on individual abundance measurements may be enough to do so, it
    is not clear what accuracy is needed to distinguish formation sites at the
    same Galactic radius.
  \item If higher accuracy is needed differential abundance analyses offer the
    possibility of reaching $\sim0.01$--$0.02$ dex accuracies as demonstrated by
    \cite{Melendez2009} and \cite{Ramirez2009} for solar analogs. Can these
    accuracies also be reached over wider ranges in the effective temperatures
    of stars? As these two papers suggest, at this level of accuracy the
    abundance patterns in stars may be affected by the presence or absence of a
    planetary system. This would then have to be accounted for in the search for
    the Sun's siblings.
\end{itemize}

With the Gaia survey starting in a few years from now, the questions above will
be actively pursued in order to ensure that precision Galactic archaeology can
be done by combining the accurate distances and kinematics from Gaia with
accurate abundances for large samples of stars throughout the Galaxy. The
results will open up the exciting prospect of further unravelling the birth
environment and life and times of the solar system through the identification of
the Sun's lost siblings.

\section*{Acknowledgements}
This work has made use of BaSTI web tools and of the SIMBAD database, operated
at CDS, Strasbourg, France. The research carried out by JB at the Sterrewacht
Leiden was made possible by the EuroScholars programme
(http://www.euroscholars.eu). We thank the anonymous referee for constructive
comments on the original manuscript.

\bibliographystyle{mn2e}
\bibliography{bpzb.bib}

\end{document}